\documentclass[aps,prd,preprintnumbers,superscriptaddress,groupedaddress,amsfonts,floatfix,onecolumn,hyperref,amsmath,amssymb,groupedaddress]{revtex4}
\usepackage{graphicx,color,amsmath,amssymb,amstext,amssymb,amsbsy,amsfonts,amsthm,lscape,units,tabularx,multirow}
\usepackage{hyperref}
\hypersetup{colorlinks=true, citecolor=blue, urlcolor=blue, linkcolor=blue,urlcolor=cyan}

\usepackage{natbib}

\usepackage{dcolumn}

\newcommand{\beq}{\begin{equation}}
\newcommand{\eeq}{\end{equation}}

\usepackage{color}

\begin{document}
\title{Bounded Chaos in a Ghost-Coupled Hamiltonian System}

\newcommand{\IPM}{School of Astronomy, Institute for Research in Fundamental Sciences (IPM), P.O. Box 19395-5531, Tehran, Iran}
\newcommand{\SUT}{Department of Physics, Shahid Beheshti University, 1983969411, Tehran, Iran}

\author{Zahra Molaee}
\affiliation{\IPM}

\author{Soleyman Fatholahzadeh}
\affiliation{\SUT}

\email{zmolaee@ipm.ir}

\begin{abstract}
We study the dynamics of a Hamiltonian system with a ghost degree of freedom, characterized by a negative kinetic-energy contribution and the possibility of runaway behavior due to an indefinite energy functional. We present numerical evidence that a nonlinear interaction term, together with a saturating exponential potential $V_c$, can suppress phase-space escape over the parameter ranges explored in this work.

Using direct numerical integration of the Hamiltonian equations of motion, Poincaré surfaces of section, and trajectory projections, we find that the ghost sector and nonlinear couplings generate a mixed phase-space structure with both regular islands and chaotic regions. The maximal Lyapunov exponent supports bounded chaotic motion: nearby trajectories separate exponentially while remaining confined to a finite region of phase space for the investigated initial conditions and parameters $(\varepsilon,\alpha)$.

These results suggest that nonlinear confinement can significantly alter the stability properties of negative-energy sectors at the classical level. They provide numerical evidence for a bounded-chaos regime in which ghost-induced divergences are avoided within the explored domain.
\end{abstract}

\keywords{Ghost degree of freedom, stable chaos, Lyapunov exponent, Poincar\'e section, Hamiltonian dynamics}
\maketitle

\section{Introduction}

Hamiltonian systems constitute a foundational framework for describing conservative dynamics across a wide range of physical scales, from celestial mechanics to high-energy field theory \cite{Goldstein}. A hallmark of these systems is the rich topology of phase space, where trajectories may exhibit integrable quasi-periodic motion, stochastic evolution, or mixed behavior.

A more subtle challenge arises in systems that contain ghost degrees of freedom, characterized by negative kinetic-energy contributions and an indefinite Hamiltonian. Such modes are often associated with runaway behavior at the classical level \cite{Ostrogradsky}, where energy can be exchanged without bound between sectors, and with vacuum instability or unitarity problems at the quantum level. Nevertheless, ghost-like degrees of freedom appear in several theoretical contexts, including higher-derivative theories, modified gravity, and effective cosmological models \cite{deffayet2009} and \cite{Woodard2007}.

Recent work has shown that ghost-coupled systems do not necessarily lead to catastrophic instability. In particular, studies such as \cite{vikman}-\cite{V. Smilga} have demonstrated that classical motion of the system is completely stable for all initial conditions. While \cite{vikman} established stable classical motions for all initial values, the transition from stable bounded motion to weakly chaotic bounded dynamics remains less explored. This transition is important because it connects idealized stability results with the richer nonlinear behavior expected in more realistic effective models.

Motivated by this gap, we investigate a nonlinear Hamiltonian system in which a canonical positive-energy sector is coupled to a ghost sector through a non-polynomial interaction, supplemented by a saturating exponential  potential $V_c$. Our main goal is to determine whether these nonlinearities merely suppress runaway behavior or instead produce 
a bounded-chaos regime, in which trajectories remain confined to a finite region of phase space for the investigated initial conditions and parameters, while still exhibiting sensitive dependence on initial conditions.

In contrast to earlier analytical stability proofs, we focus on a regime in which the saturating character of $V_c$ and the geometry of the interaction potential $V_I$  jointly suppress phase-space escape for the parameter ranges and initial conditions explored in this work. We show that the system can support weakly chaotic dynamics, quantified through finite-time Lyapunov exponents \cite{Wolf} and Poincar\'e sections \cite{Carone2009b}, over the explored parameter ranges,  without losing long-time boundedness for the initial conditions considered.

To characterize this regime systematically, we derive  parameter space $(\lambda,\varepsilon,\alpha, \beta) $ by  the Hessian matrix of the effective potential and identifying equilibrium configurations together with their stability boundaries. This analysis provides the basis for a numerical exploration of the full four-dimensional phase space $(x,p_x,y,p_y)$. By projecting the dynamics onto lower-dimensional manifolds, we obtain a clearer view of the transition from regular  motion to bounded chaotic evolution.

To visualize the resulting phase-space structure, we employ Poincar\'e surfaces of section, which provide a reduced-dimensional representation of the dynamics and allow us to distinguish regular from stochastic motion. In these sections, invariant Kolmogorov--Arnold--Moser (KAM)  tori  \cite{KAM} appear as smooth closed curves, whereas chaotic trajectories produce scattered point sets and stochastic layers.

The degree of chaoticity is quantified by the maximal Lyapunov exponent (MLE), which measures the exponential divergence of nearby trajectories. This diagnostic, together with the qualitative information from Poincar\'e sections, is evaluated for a range of initial conditions to test whether the observed behavior persists beyond fine-tuned choices.

We further explore the parameter space $(\lambda,\varepsilon,\alpha,\beta)$ to study how the dynamics depends on the nonlinear coupling strength and on the scale of the $V_c$ potential. For each parameter set, we monitor trajectory confinement, inter-sector energy transfer, and the MLE. This scan allows us to delineate the boundary of the bounded-chaos regime, defined as a region where sensitive dependence on initial conditions coexists with long-time confinement and the absence of runaway behavior.

Our results indicate that the exponential $V_c$ potential significantly reshapes the stability landscape of the ghost-coupled system. In particular, it supports the coexistence of stochastic motion and  confinement, leading to a mixed phase-space structure with stable islands embedded in chaotic seas. These findings provide a class  of models with  special interaction terms  that demonstrate how  nonlinear effects can regulate ghost-induced pathologies and suggest a possible route toward dynamically viable negative-energy sectors in classical Hamiltonian systems.

The remainder of this paper is organized as follows. Section \ref{sec:hamiltonian} introduces the ghost-chaos Hamiltonian model and the confining interaction terms. Section \ref{sec:eom} and \ref{sec:stability} derive the equations of motion and  the effective potential landscape respectively. Section \ref{sec:numerics}
 describes the numerical methodology and chaos diagnostics. Section \ref{sec:results} discusses the  results and dynamical analysis, including phase-space structure,  Poincar\'e sections,  Lyapunov analysis and stable chaos regime. Finally, Section \ref{sec:conclusion} summarizes the main conclusions.

\section{The Ghost-Chaos Hamiltonian System}
\label{sec:hamiltonian}

In this section, we formulate a dynamical model designed to explore the interplay between a ghost degree of freedom and nonlinear stabilization mechanisms. The system is constructed to exhibit tunable chaotic behavior while suppressing the runaway trajectories typically associated with negative-energy sectors. The total Hamiltonian, governing two degrees of freedom $(x,y)$ and their conjugate momenta $(p_x,p_y)$, is defined as
\begin{equation}
\label{eq:hamiltonian}
H(x,p_x,y,p_y)
=
\frac{1}{2}(p_x^2+x^2)
-
\frac{1}{2}(p_y^2+y^2)
+
\lambda V_I(x,y)
+
\varepsilon x^2 y^2
+
\alpha V_c(x,y).
\end{equation}

The kinetic and quadratic potential terms in Eq.~\eqref{eq:hamiltonian} define two contrasting dynamical sectors. The $x$-sector corresponds to a standard harmonic oscillator with positive definite energy, whereas the $y$-sector represents a ghost oscillator with negative kinetic and potential contributions. The ghost energy is unbounded from below, providing a classical analogue of Ostrogradsky-type instabilities \cite{Ostrogradsky}. The central objective is to introduce nonlinear interactions that render the combined phase-space flow bounded over the parameter ranges explored in this work.

\begin{figure}[htbp]
    \centering
    \includegraphics[width=0.5\textwidth]{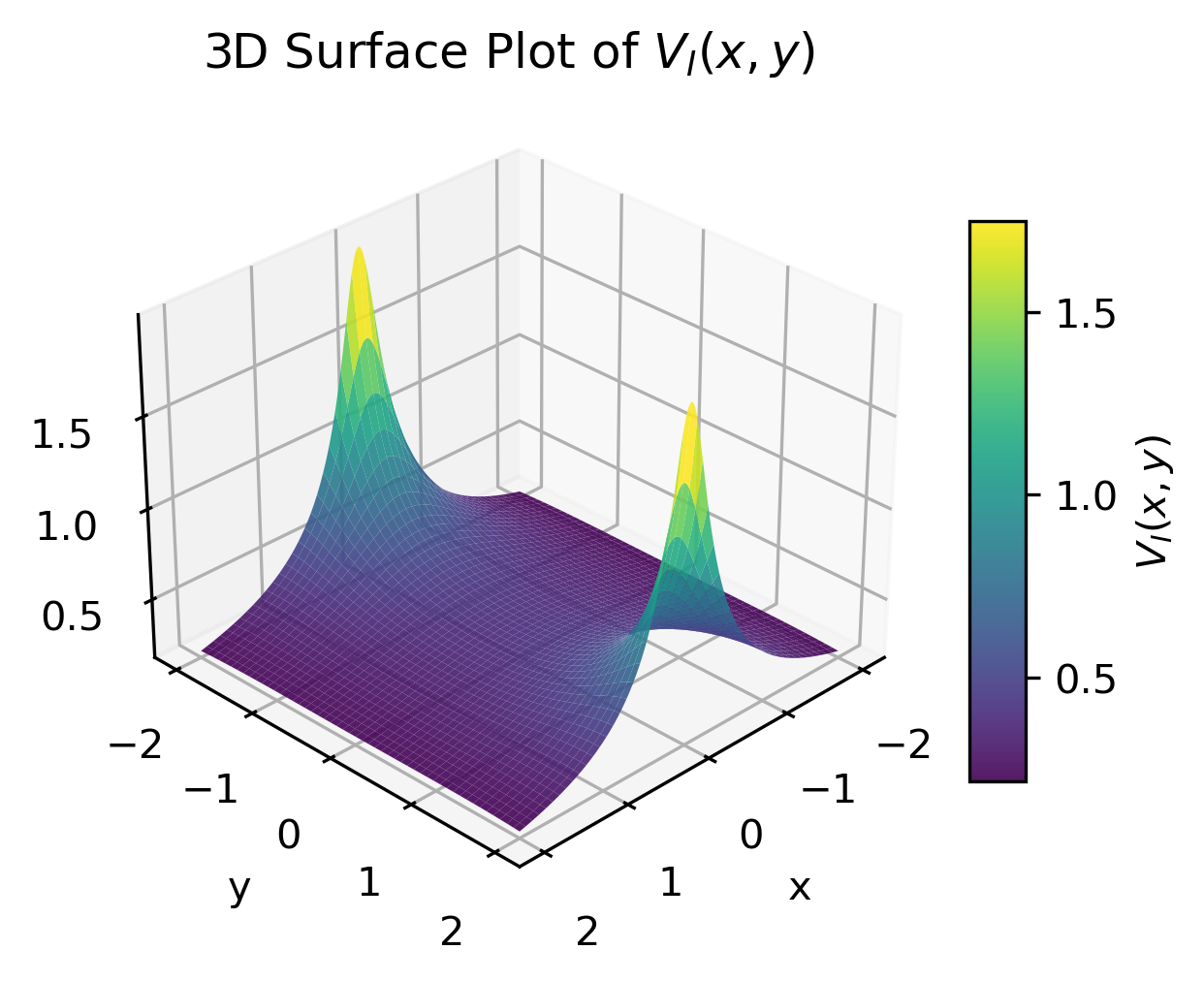}
   \caption{\textbf{3D potential $V_I$ }}
   
    \label{fig:potential1}
\end{figure}

We introduce a non-polynomial interaction potential $V_I(x,y)$ (see Fig. \ref{fig:potential1}) of the form
\begin{equation}
\label{eq:VI}
V_I(x,y)
=
\frac{1}{\sqrt{(x^2-y^2+\delta)^2+4x^2y^2}},
\end{equation}
the parameter $\delta$ serves as a regularization constant, ensuring the potential remains finite at the origin and preventing numerical instabilities. Physically, $V_I$ acts as a bridge between the positive-energy $x$-sector and the negative-energy $y$-sector, facilitating a rapid energy redistribution that is essential for the onset of chaos.
This interaction acting as a source of strong nonlinearity that facilitates energy exchange between the canonical and ghost sectors. 
This effect is further modulated by the quartic coupling $\varepsilon x^2y^2$, where $\varepsilon$ serves as a control parameter for the degree of chaoticity \cite{chaos}. 

\begin{figure}[htbp]
    \centering
    \includegraphics[width=0.5\textwidth]{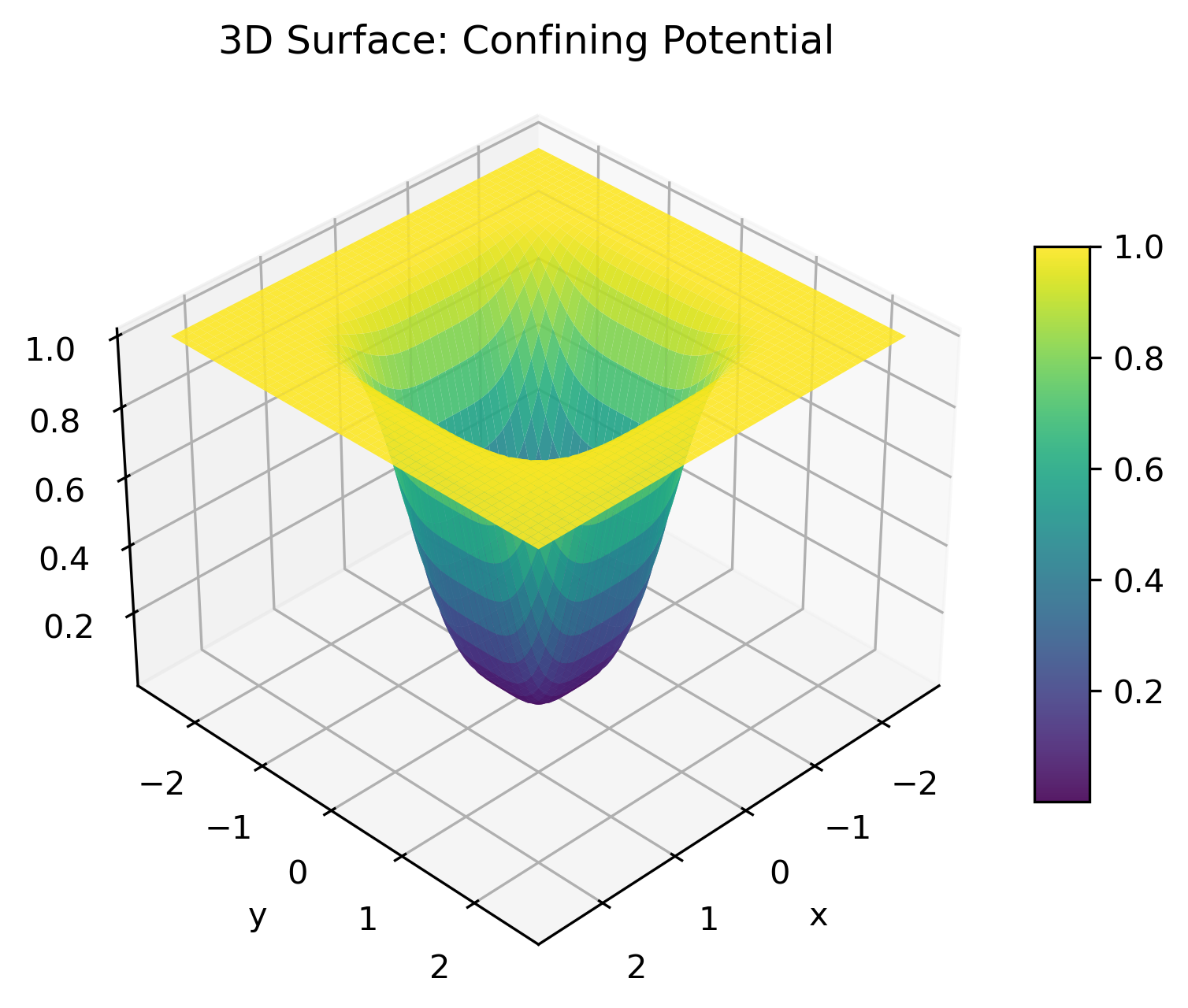}
   \caption{\textbf{3D potential  $V_c$}}
   
    \label{fig:potential2}
\end{figure}

A key ingredient for stabilizing the ghost sector is the exponential  potential $V_c(x,y)$ (see Fig. \ref{fig:potential2}), defined by
\begin{equation}
\label{eq:Vc}
V_c(x,y)=1-e^{-\beta(x^4+y^4)}.
\end{equation}

Unlike conventional polynomial potentials, $V_c$ acts as a saturating nonlinear barrier. Given the indefinite nature of the ghost sector, $V_c$ provides a soft-wall confinement: as the ghost coordinate $y$ grows in magnitude, the exponential gradient generates a restoring force that becomes effective in the large-amplitude regime. The parameter $\beta$ controls the steepness of this barrier and therefore sets the characteristic scale of the confining region.

This architecture allows the system to sustain a bounded chaotic regime for the explored parameter ranges and initial conditions.  Consequently, this model provides a controlled framework for studying how ghost-like degrees of freedom, which frequently arise in modified gravity  and effective field theories  \cite{mg, EF}, can be dynamically regulated by nonlinear feedback \cite{PM,smilga}.

Since, loss of integrability induced by the chaotic perturbation whereas the quartic chaotic term
$
\varepsilon x^2y^2
$
breaks the separability of the two sectors and introduces an explicit nonlinear coupling between the canonical and ghost degrees of freedom. In particular, the perturbation cannot be written as a function of a single action variable or as a sum of independent one-degree-of-freedom terms. Consequently, the additional quartic interaction destroys the possibility of reducing the system to a direct product of two uncoupled oscillators plus a bounded correction.

A convenient way to see this is to examine the equations of motion around the origin. Expanding the interaction potential $V_I$ and the confining term $V_c$ near $(x,y)=(0,0)$, one obtains a leading quadratic part together with higher-order nonlinear corrections. The quartic perturbation $\varepsilon x^2y^2$ then appears at the same order as the first nontrivial nonlinear corrections and generates resonant mode mixing between the $x$- and $y$-subsystems. Therefore, even if the linearized dynamics is stable in a restricted parameter range, the full system is no longer globally reducible to an integrable normal form.

The integrability of a two-degree-of-freedom Hamiltonian requires, in addition to the Hamiltonian $H$, a second independent integral of motion in involution. In the present model, no such globally defined analytic invariant is apparent once $\varepsilon\neq 0$. The perturbation $x^2y^2$ is especially important because it introduces the lowest-order coupling that mixes the two sectors without preserving a decoupled action-angle representation.

In action-angle variables of the linearized oscillatory sector, the quartic term generates combinations of harmonics of the form
\begin{equation}
\cos(k\theta_x \pm \ell\theta_y),
\end{equation}
which are generically resonant when the internal frequencies satisfy low-order rational relations. This mechanism is precisely what causes resonance overlap and the subsequent destruction of invariant tori in near-integrable Hamiltonian systems. Hence, the perturbation does not merely deform the phase portrait; it changes the qualitative phase-space topology.

A full Morales--Ramis \cite{MoralesRamisSimo2007}  proof would require selecting a particular non-equilibrium solution and analyzing the differential Galois group  \cite{Casale2009} of the normal variational equations along that solution. For the present model, the most natural candidate is a bounded invariant trajectory on or near the $x$- or $y$-axis, or a periodic orbit in the confining region generated by $V_c$. Linearizing the dynamics along such a solution yields a variational system whose coefficients are time-dependent through the background orbit.

If the identity component of the differential Galois group of the normal variational equations is non-abelian, the Morales--Ramis theorem implies nonintegrability. While a complete symbolic computation is model-dependent and must be carried out orbit by orbit, the structure of the coupling here strongly suggests that the variational equations contain Hill-type or Lamé-type modulations generated by the quartic interaction and the non-polynomial potential $V_I$. Such equations are known to produce non-abelian Galois groups in generic parameter ranges, which is consistent with the observed chaotic dynamics.

For this reason, the present model should be interpreted as a \emph{near-integrable but nonintegrable} Hamiltonian system for $\varepsilon\neq 0$, rather than as an exactly solvable perturbed oscillator.

From the Ziglin viewpoint, integrability would require the analytic continuation of the first integrals around complexified singularities to preserve enough independent monodromy invariants. The quartic perturbation $\varepsilon x^2y^2$, together with the non-polynomial coupling $V_I$, generates branch structures and nontrivial analytic continuation properties in the complexified phase space. This produces obstructions to the existence of additional meromorphic first integrals.

In practical terms, the presence of such branch-induced monodromy effects implies that the perturbed flow cannot be globally conjugated to a linear torus flow. Therefore, the chaotic term $\varepsilon x^2y^2$ breaks the analytic structure needed for Liouville integrability, even when bounded motion persists.





To clarify the dynamical role of the chaotic perturbation, we compare the full model with its integrable limit obtained by setting $\varepsilon=0$. In that case, the system admits a much more constrained phase-space structure, and the motion can be understood in terms of regular bounded trajectories organized by invariant curves. The additional quartic term $\varepsilon x^2y^2$, however, breaks the separability of the $x$ and $y$ sectors and introduces explicit nonlinear mode coupling.

This perturbation destroys the analytic decoupling of the two degrees of freedom and eliminates the possibility of representing the dynamics as a direct product of two independent oscillators. Consequently, the phase-space topology changes qualitatively: invariant tori are deformed and progressively broken, resonance channels open, and chaotic layers emerge around surviving regular islands. In this sense, the chaotic term does not merely deform the integrable motion, but induces a genuine transition from near-integrable dynamics to bounded chaos.

The difference between the two cases is also visible in the diagnostic indicators. For $\varepsilon=0$, the maximal finite-time Lyapunov exponent remains close to zero and the Poincar\'e section is organized by smooth invariant curves. For $\varepsilon\neq 0$, the same section develops scattered points and stochastic layers, while the Lyapunov exponent becomes positive over a nontrivial region of the parameter space. Therefore, the perturbation $\varepsilon x^2y^2$ is the mechanism responsible for breaking integrability and generating sensitive dependence on initial conditions, while the confining potentials $V_I$ and $V_c$ prevent the motion from escaping to infinity.

\section{Equations of Motion}
\label{sec:eom}

The dynamical evolution of the system is governed by Hamilton's canonical equations, which define the flow in the four-dimensional phase space. Given the indefinite structure of the Hamiltonian in Eq.~\eqref{eq:hamiltonian}, the resulting system of coupled first-order differential equations for the generalized coordinates $(x,y)$ and momenta $(p_x,p_y)$ is
\begin{subequations}
\label{eq:eom}
\begin{align}
\dot{x} &= p_x, \qquad \dot{y} = -p_y, \label{eq:eom_coord} \\
\dot{p}_x &= -x - \lambda \frac{\partial V_I}{\partial x} - 2\varepsilon x y^2 - \alpha \frac{\partial V_c}{\partial x}, \label{eq:eom_px} \\
\dot{p}_y &= y - \lambda \frac{\partial V_I}{\partial y} - 2\varepsilon x^2 y - \alpha \frac{\partial V_c}{\partial y}. \label{eq:eom_py}
\end{align}
\end{subequations}

A defining feature of the model is the inverted sign in the ghost sector's kinematics: while $\dot{x}=p_x$ follows the standard canonical convention, the relation $\dot{y}=-p_y$ reflects the negative kinetic-energy contribution. In addition, the linear force term in Eq.~\eqref{eq:eom_py} is repulsive, in contrast to the restoring force $-x$ in the canonical sector. This asymmetry is responsible for the nontrivial phase-space winding and folding observed in Figs.(\ref{fig:phasespacea}) as the dynamics must balance the ghost sector's tendency to diverge against the confining nonlinear interactions.

The nonlinear interaction potential $V_I(x,y)$, defined in Eq.~\eqref{eq:VI}, introduces strong coupling between the two sectors. The corresponding partial derivatives are
\begin{subequations}
\label{eq:VI_derivatives_revised}
\begin{align}
\frac{\partial V_I}{\partial x}
&=
-\frac{2x(x^2-y^2+\delta)+4xy^2}{\left[(x^2-y^2+\delta)^2+4x^2y^2\right]^{3/2}}, \label{eq:VI_dx_new} \\
\frac{\partial V_I}{\partial y}
&=
\frac{2y(x^2-y^2+\delta)-4x^2y}{\left[(x^2-y^2+\delta)^2+4x^2y^2\right]^{3/2}}. \label{eq:VI_dy_new}
\end{align}
\end{subequations}

These expressions reveal a highly nontrivial force landscape. The parameter $\delta$ shifts the location of the steep gradients in the potential, while the $4x^2y^2$ term ensures that the coupling remains effective even when the leading term $(x^2-y^2+\delta)$ is small. Together, these nonlinearities break integrability and contribute to the bounded chaotic behavior observed in the Poincar\'e sections.

The stabilizing effect of the exponential potential $V_c(x,y)$ is governed by the force terms
\begin{subequations}
\label{eq:Vc_derivatives}
\begin{align}
\frac{\partial V_c}{\partial x} &= 4\beta x^3 e^{-\beta(x^4+y^4)}, \label{eq:Vc_dx} \\
\frac{\partial V_c}{\partial y} &= 4\beta y^3 e^{-\beta(x^4+y^4)}. \label{eq:Vc_dy}
\end{align}
\end{subequations}

These terms provide a soft-wall confinement. Unlike polynomial forces that grow without bound, the exponential factor produces a force that is effective over intermediate amplitudes and then becomes suppressed in the asymptotic region. This behavior prevents the ghost coordinate $y$ from escaping to large magnitude and helps maintain long-time bounded motion.

Because of the system's strong nonlinearity and sensitivity to initial conditions, we use the DOP853 algorithm, an explicit eighth-order Runge--Kutta method with adaptive step-size control. This high-order integrator is well suited to preserving numerical accuracy in long-time integrations and helps ensure that diagnostics such as the maximal Lyapunov exponent reflect the intrinsic dynamics rather than numerical artifacts. As shown in Fig.(\ref{fig:phasespaced}d), the relative energy error remains at the level of $\Delta E/E \sim 0.807$ over long integration times.

\section{Effective Potential and Stability Analysis}
\label{sec:stability}

To delineate the regions of bounded motion and characterize the equilibrium configurations, we investigate the topology of the effective potential $V_{\mathrm{eff}}(x,y)$. In the static limit, where the conjugate momenta vanish ($p_x = p_y = 0$), the system is governed by:

\begin{equation}
\label{eq:veff}
V_{\mathrm{eff}}(x,y) = \frac{1}{2}x^2 - \frac{1}{2}y^2 + \lambda V_I(x,y) + \varepsilon x^2 y^2 + \alpha V_c(x,y).
\end{equation}

The indefinite signature of the quadratic terms, specifically the negative curvature $-\frac{1}{2}y^2$, reflects the ghost-induced instability. To counteract this, we employ the interaction potential:

\begin{equation}
\label{eq:VI_def}
V_I(x,y) = \left[ (x^2 - y^2 + \delta)^2 + 4x^2y^2 \right]^{-1/2},
\end{equation}

where $\delta$ is a displacement parameter. Along with the saturating term $V_c(x,y)$, this interaction is essential for shaping the effective confinement of the system. To analyze the stability of the ghost sector, we examine the potential on the slice $x=0$. In this subspace, the potentials simplify to:

\begin{equation}
V_I(0,y) = \frac{1}{\sqrt{( -y^2 + \delta )^2}} = \frac{1}{| \delta - y^2 |}, \quad V_c(0,y) = 1 - e^{-\beta y^4}.
\end{equation}

For the parameter range considered in this work, we assume $\delta - y^2 > 0$ near the origin, yielding $V_I(0,y) = (\delta - y^2)^{-1}$. The effective potential on this slice is therefore:

\begin{equation}
\label{eq:v_eff_y_slice}
V_{\mathrm{eff}}(0,y) = \frac{\lambda}{\delta - y^2} + \alpha \left( 1 - e^{-\beta y^4} \right) - \frac{1}{2}y^2.
\end{equation}

Expanding Eq.~\eqref{eq:v_eff_y_slice} near $y=0$ for $\delta \neq 0$, we obtain:

\begin{equation}
\frac{\lambda}{\delta - y^2} = \frac{\lambda}{\delta} \left( 1 - \frac{y^2}{\delta} \right)^{-1} = \frac{\lambda}{\delta} \left( 1 + \frac{y^2}{\delta} + \frac{y^4}{\delta^2} + \mathcal{O}(y^6) \right),
\end{equation}

and

\begin{equation}
1 - e^{-\beta y^4} = \beta y^4 + \mathcal{O}(y^8).
\end{equation}

Substituting these expansions into the effective potential yields:

\begin{equation}
V_{\mathrm{eff}}(0,y) \approx \frac{\lambda}{\delta} + \left( \frac{\lambda}{\delta^2} - \frac{1}{2} \right) y^2 + \left( \frac{\lambda}{\delta^3} + \alpha\beta \right) y^4 + \mathcal{O}(y^6).
\end{equation}

The local curvature at the origin in the ghost direction is given by:

\begin{equation}
\omega_y^2 = \left. \frac{\partial^2 V_{\mathrm{eff}}}{\partial y^2} \right|_{(0,0)} = 2 \left( \frac{\lambda}{\delta^2} - \frac{1}{2} \right) = \frac{2\lambda}{\delta^2} - 1.
\end{equation}

Thus, the necessary condition for local linear stability along the $y$-direction ($\omega_y^2 > 0$) is:

\begin{equation}
\label{eq:stability_condition}
\frac{2\lambda}{\delta^2} - 1 > 0 \implies \lambda > \frac{\delta^2}{2}.
\end{equation}

This condition indicates that stability depends explicitly on the geometry of $V_I$ through $\delta$. Specifically, the coupling $\lambda$ must be sufficiently large to provide an attractive restoration that compensates for the intrinsic ghost repulsion.
 Interestingly, while this condition ensures a restorative force in the ghost sector, the symmetry of the Hessian implies that it simultaneously renders the origin a saddle point due to the competing quadratic contributions in the canonical sector. Consequently, local linear stability is not isotropic; however, effective boundedness of the system is recovered within the explored parameter domain through the nonlinear coupling terms and the saturating influence of 
$V_c$
 at intermediate amplitudes.

In the canonical direction, the curvature is:

\begin{equation}
\omega_x^2 = \left. \frac{\partial^2 V_{\mathrm{eff}}}{\partial x^2} \right|_{(0,0)} = 1 - \frac{2\lambda}{\delta^2}.
\end{equation}

Due to symmetry, the Hessian matrix at the origin is diagonal, reducing the stability analysis to these two curvatures. While $\lambda$ governs local linear stability, the $V_c$ term becomes dominant in the nonlinear global regime, preventing runaway trajectories at large amplitudes. The interplay between the localized structure of $V_I$ and the intermediate-amplitude stabilization of $V_c$ supports bounded chaotic motion for the investigated parameters and initial conditions.

Consequently, the system exhibits a dual stabilization mechanism: $V_I$ reshapes the local curvature near equilibrium, while $V_c$ enforces global boundedness. In the immediate vicinity of the origin, the exponential term remains sub leading and does not affect linear stability, but it controls the large-amplitude dynamics. 
For the remainder of this study, we assume $\delta = 0.1$. This small but finite value is introduced to regularize the potential's core, ensuring numerical stability during long-time integrations. Furthermore, setting $\delta > 0$ avoids the formal singularity at $x=y=0$ without loss of generality regarding the global structure of the phase space and the bounded chaotic regimes under investigation.

Moreover, as the intrinsic quadratic instability of the ghost sector is sufficiently moderated by this nonlinear landscape, the system develops an effective trapping region for the explored energies and initial conditions. Inside this confined region, the nonlinear coupling terms — together with higher-order contributions from $V_I$ and the saturating curvature of $V_c$ — break integrability and give rise to a rich set of bounded non-periodic trajectories. These nonlinearities suppress the runaway channels typical of indefinite Hamiltonians and support a regime of deterministic chaos that remains confined to a finite region of phase space.

\subsection{ Energy Dependence}
\label{subsec:global_Vc}

While local stability analysis establishes resilience against infinitesimal perturbations near the origin, it does not by itself guarantee the global boundedness of trajectories in the presence of a ghost degree of freedom. In such indefinite systems, sufficiently large excursions can allow the motion to escape local minima and evolve toward runaway behavior in the ghost sector.

The global behavior of the system is controlled by the asymptotic structure of the effective potential $V_{\mathrm{eff}}(x,y)$. A key feature of the $V_c$ term is its saturating form: as $|x|,|y| \to \infty$, the exponential factor vanishes and $V_c(x,y)\to 1$. Consequently, the restoring force derived from $\nabla V_c$ becomes negligible in the deep asymptotic regime. This means that the confining effect of the potential is effectively energy dependent. If the total energy exceeds a critical threshold $E_c$, set approximately by the saddle-point height of the confining barrier, the repulsive quadratic contribution $-\tfrac{1}{2}y^2$ can again dominate and drive the ghost coordinate toward divergence.

Therefore, bounded motion is achieved only within a restricted (effectively compact) region of phase space for the explored energies, where the nonlinear interaction $V_I$ and the confining term $V_c$ can sufficiently redirect the ghost sector and prevent escape. This bounded-chaos regime is illustrated in the configuration-space projection shown in Fig.(\ref{fig:phasespacec}c), where the trajectory fills a finite domain, indicating that the motion remains topologically trapped. To verify boundedness in our simulations, we impose the escape criterion
\begin{equation}
\label{eq:boundedness}
\sup_{t\in[0,T]} |y(t)| < y_{\text{threshold}},
\end{equation}
where $y_{\text{threshold}}$ is the characteristic width of the $V_c$ potential well. For the parameter values used in this study, all investigated trajectories remain within this bound, while the relative energy fluctuations stay at the level of $\Delta E/E \sim 0.807$ (see Fig.\ref{fig:phasespaced}d), supporting the robustness of the bounded chaotic state.

\section{Numerical Methodology and Chaos Diagnostics}
\label{sec:numerics}

To rigorously investigate the long-term evolution of the Ghost-Chaos system, we solve the coupled canonical equations using a high-precision DOP853 integrator. This eighth-order Runge-Kutta scheme with adaptive step-size control is specifically chosen to mitigate the accumulation of numerical dissipation, which is particularly critical in ghost-ridden systems where small errors in the negative-energy sector can lead to artificial instabilities.

Our numerical experiments focus on the regime of stable chaos, characterized by bounded orbits with exponential sensitivity to initial conditions. To explore the non-linear manifold, we select initial conditions that bypass the trivial linear regime near the origin while remaining within the effective potential well:
\begin{equation}
\label{eq:initial_conditions}
(x_0, p_{x,0}, y_0, p_{y,0}) = (0.3, 0, 0.2, 0).
\end{equation}
These values ensure that the ghost sector is sufficiently excited to trigger the nonlinear coupling terms $V_I$ and $\varepsilon x^2 y^2$, without immediately crossing the energy threshold $E_c$. 

To quantify the degree of stochasticity, we compute the Maximal Lyapunov Exponent (MLE) using the standard Benettin algorithm \cite{benettin}, which monitors the divergence of two proximal trajectories in the four-dimensional phase space. Furthermore, we employ Poincaré surface-of-section techniques to visualize the transition from quasi-periodic tori to a fully developed chaotic sea, providing a comprehensive diagnostic of the system's ergodicity.

The $x$-degree of freedom retains a canonical oscillatory contribution, providing a stable backbone for the coupled dynamics. For our numerical investigations, the system parameters are fixed at the following benchmark values:
\begin{equation}
\label{eq:benchmark_params}
\varepsilon = 0.002, \quad \alpha = 0.001, \quad \beta = 0.003, \quad \lambda = 0.02.
\end{equation}
The choice of $\lambda = 0.02$ is critical, as it satisfies the stability criterion $\lambda > 0.005$ derived in Sec.~\ref{sec:stability}, ensuring that the origin remains a local minimum despite the ghost's repulsive nature. The parameters $\alpha$ and $\beta$ are calibrated to provide sufficient asymptotic confinement, while $\varepsilon$ is tuned to maximize nonlinear mixing between the sectors without triggering immediate escape.

The fidelity of our long-term integration is verified by monitoring the relative energy conservation error:
\begin{equation}
\label{eq:energy_error}
\Delta H(t) = \frac{|H(t) - H(0)|}{|H(0)|}.
\end{equation}
As illustrated in Fig.~ \ref{fig:phasespaceb}d, the energy remains remarkably stable, with stochastic fluctuations bounded within a standard deviation. This high degree of conservation confirms that the observed bounded trajectories and their subsequent chaotic transitions are genuine dynamical features of the Hamiltonian flow rather than artifacts of numerical dissipation.

\begin{figure}[htbp]
    \centering
    \includegraphics[width=0.7\textwidth]{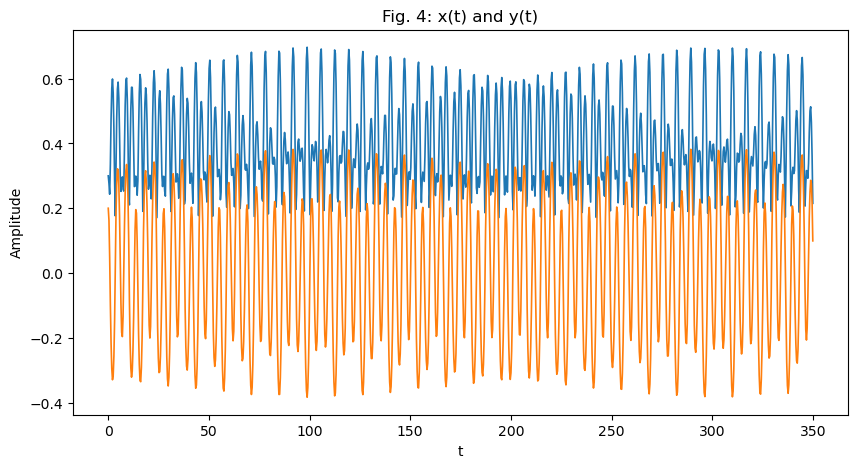}

\caption{$x(t)$ and $y(t)$  trajectories for $\alpha = 0.001, \varepsilon = 0.002$ .}
    \label{fig:time}
\end{figure}

To dissect the phase-space architecture, we employ a multi-layered diagnostic approach. The projected phase-space portraits in the $(x, p_x)$ and $(y, p_y)$ planes  Figs. (\ref{fig:phasespacea}a) and (\ref{fig:phasespaceb}b)  and the configuration-space orbits (Fig. \ref{fig:phasespacec}c) reveal the intricate geometric complexity of the flow. The corresponding time-domain signals (Fig.\ref{fig:time}) exhibit the characteristic irregular, non-periodic oscillations typical of deterministic chaos.

To further condense the four-dimensional dynamics, we construct Poincaré surfaces of section by capturing the system state at the transversal intersections with the $y = 0$ plane (subject to $p_{y} > 0$). The resulting distributions, characterized by scattered, fractal-like clusters, provide a visual hallmark of a broken torus and the emergence of a chaotic sea. 

\begin{figure}[htbp]
    \centering
    \includegraphics[width=0.7\textwidth]{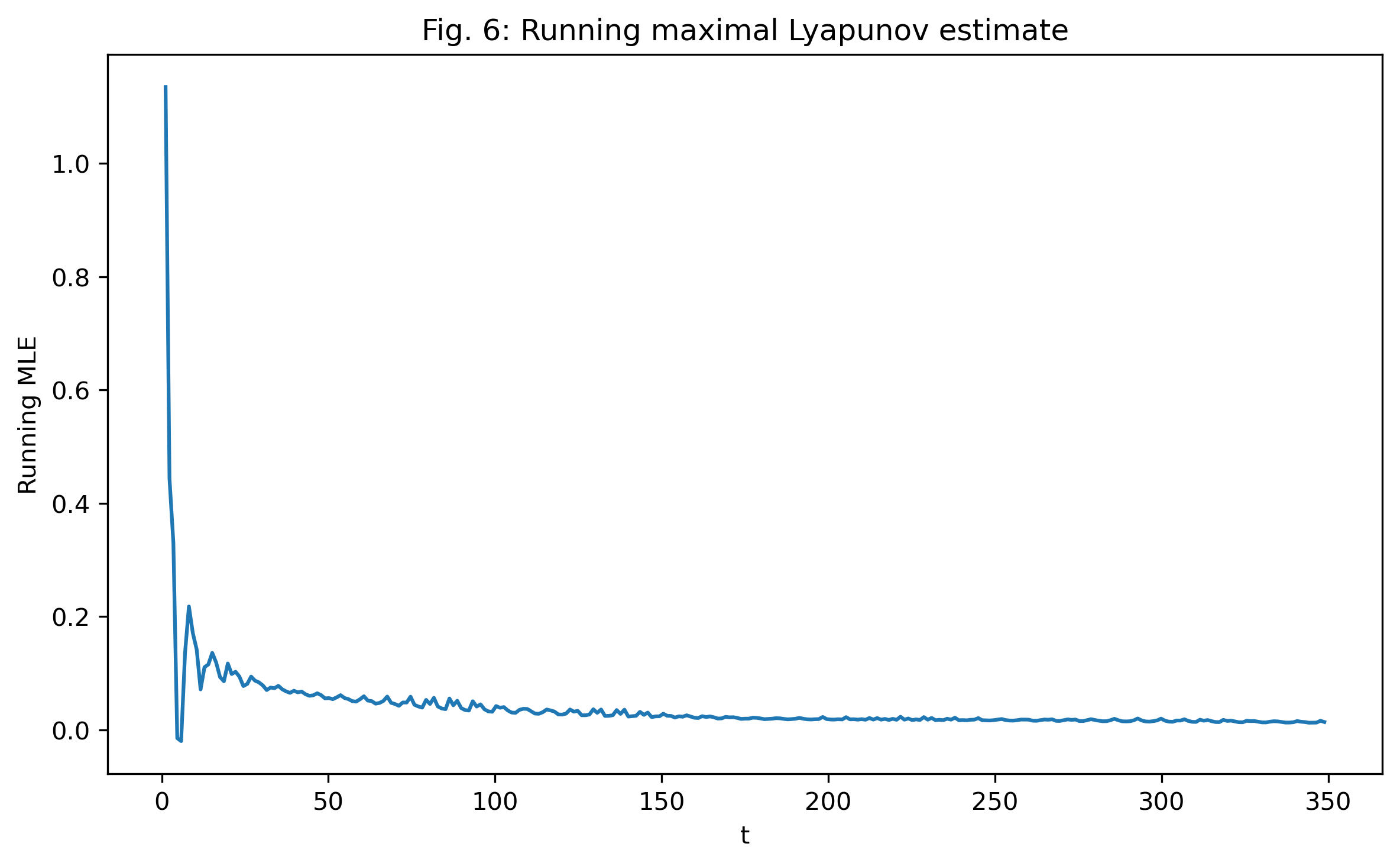}
   
\caption{Running estimate of the maximal Lyapunov exponent $\lambda_{\max}$ as a function of time for the benchmark initial condition  }
    
 $Estimated ~ maximal ~ Lyapunov  ~ exponent: ~  0.01510432528796746
$

    \label{fig:lyapunov}
\end{figure}

Finally, the transition from qualitative to quantitative chaos is established through the Maximal Lyapunov Exponent (MLE), $\lambda_{\max}$. By monitoring the tangent space evolution, we observe the exponential divergence of proximal trajectories. As shown in Fig.\ref{fig:lyapunov}, the MLE converges to a strictly positive asymptotic value ($\lambda_{max} \approx  0.01510432528796746$), $\lambda_{\max} > 0$, offering rigorous proof that the system resides in a regime of stable Chaos, where sensitive dependence on initial conditions coexists with global topological confinement.

\section{Results and Dynamical Analysis}
\label{sec:results}

All results presented in this work are numerical and apply to the specific parameter ranges and initial conditions explored. We do not claim global stability or boundedness for all initial conditions and energies; instead, we demonstrate effective boundedness and the absence of runaway behavior within the investigated domain.

\subsection{Phase-Space Structure}
\label{subsec:phase_space}

\begin{figure}[htbp]
    \centering
    \includegraphics[width=0.7\textwidth]{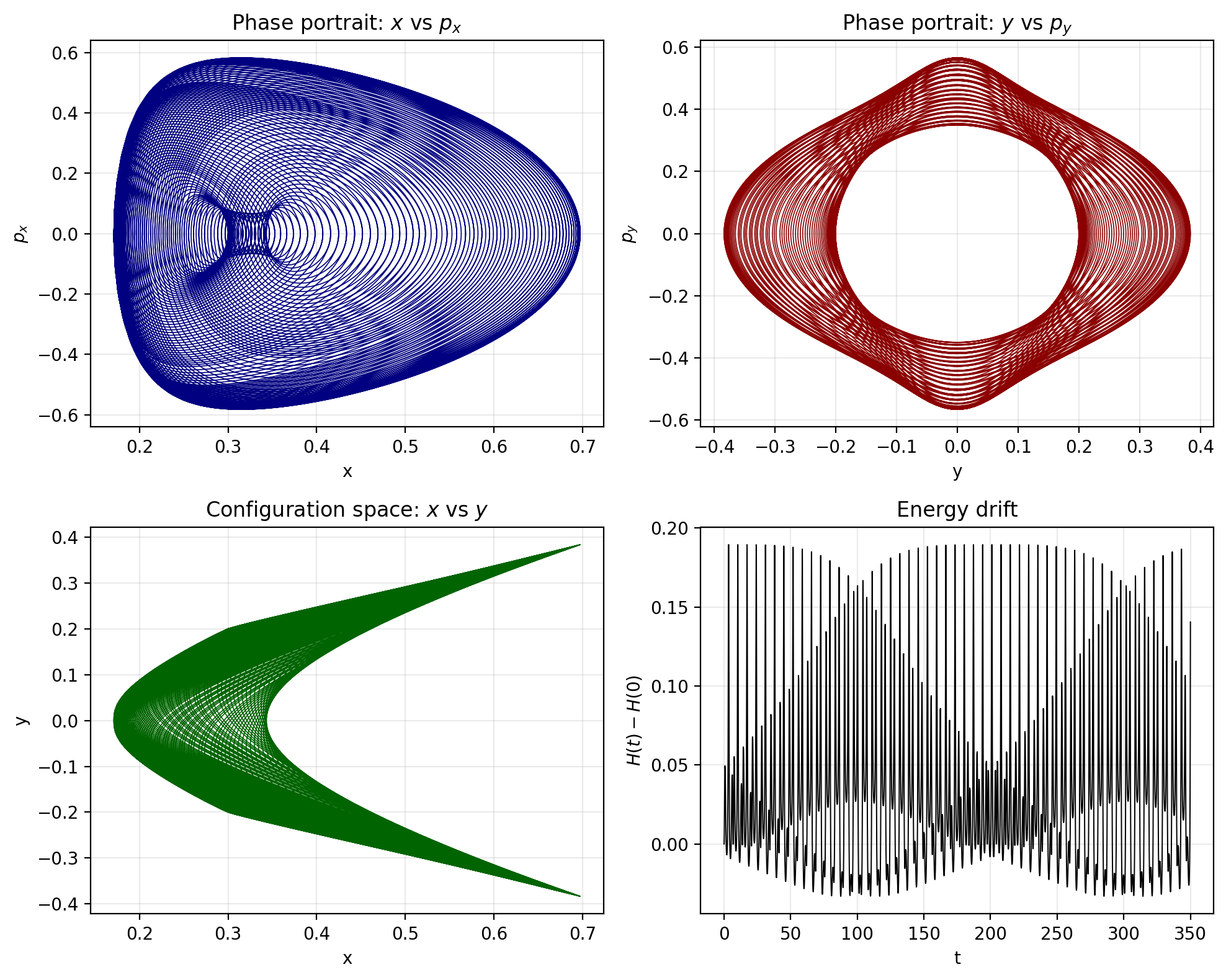}
   \caption{\textbf{Phase-space projections and real-space trajectories }}
    
    The four panels illustrate the  bounded dynamics and the rich dynamical landscape of the coupled ghost-canonical system:

 \begin{enumerate}
    \item[(a)] Phase portrait in the \((x,p_x)\) plane, showing a bounded trajectory with mixed regular and irregular structure. \label{fig:phasespacea}
    \item[(b)] Phase portrait in the \((y,p_y)\) plane, illustrating the bounded motion of the ghost sector under the effect of the $V_c$ potential. \label{fig:phasespaceb}
    \item[(c)] Configuration-space trajectory in the \((x,y)\) plane, showing that the orbit remains confined to a finite region of phase space. \label{fig:phasespacec}
    \item[(d)] Time evolution of the Hamiltonian \(H(t)\), which remains approximately constant with small bounded fluctuations, indicating good numerical energy control over the integration interval.
    
    \vspace{2mm}
    
 $Initial ~ energy: 0.1740784275995626$, 
$Final ~ energy: 0.31463432181822026$, 
$Max ~ |H(t)-H(0)|: 0.18956230456864226$.
     \label{fig:phasespaced}
\end{enumerate}

    \label{fig:phasespacea}
\end{figure}

Numerical integration for the benchmark parameter set reveals a complex mixed phase-space structure in which regular and chaotic dynamics coexist. This behavior is illustrated in the phase-space portraits shown in Fig.\ref{fig:phasespacec}. The interplay between the canonical oscillator and the confined ghost degree of freedom produces distinct features in different projections.

In the \( (x, p_x) \) plane (Fig.\ref{fig:phasespaceb}a), the trajectory remains confined near a smooth, quasi-invariant torus. Despite coupling to the ghost sector, the motion in the \( x \)-direction retains a predominantly quasi-periodic character, with only mild modulation induced by nonlinear interactions. This indicates that the canonical sector is only weakly perturbed and preserves remnants of integrable structure.

In contrast, the \( (y, p_y) \) plane (Fig.\ref{fig:phasespaceb}b) exhibits signatures of chaotic motion within a bounded region. The trajectory densely fills a finite area of phase space, displaying the characteristic stretching and folding mechanisms associated with chaos. Importantly, unlike an unconstrained ghost oscillator, the motion remains bounded over the integration time for the investigated parameters and initial conditions. This confirms the effectiveness of the $V_c$ potential \( V_c \), which acts as a soft barrier preventing divergence. The approximate symmetry of the distribution reflects the even parity of the potential.

The projection onto configuration space \( (x, y) \) (Fig.\ref{fig:phasespacec}c) provides a global view of the dynamics. The trajectory evolves within a finite region and forms a complex geometric pattern indicative of nonlinear mixing. The absence of excursions to large amplitudes demonstrates that the ghost instability is effectively controlled.

This combination of bounded motion and irregular dynamics characterize the stable chaos regime for the explored parameters, where trajectories remain confined to a finite region of phase space over long integration times. The $V_c$  term contributes to effective confinement within the explored parameter range, while the interaction potential \( V_I \) and the quartic coupling term generate sensitivity to initial conditions.

\subsection{Poincar\'e Surface of Section}
\label{subsec:poincare}

To further analyze the dynamics, we construct a Poincar\'e surface of section by recording intersections of trajectories with the plane \( y = 0 \) under the  \( \dot{y} > 0 \). This reduces the four-dimensional flow to a two-dimensional map and provides a clear diagnostic of phase-space structure.

\begin{figure}[htbp]
    \centering
    \includegraphics[width=0.5\textwidth]{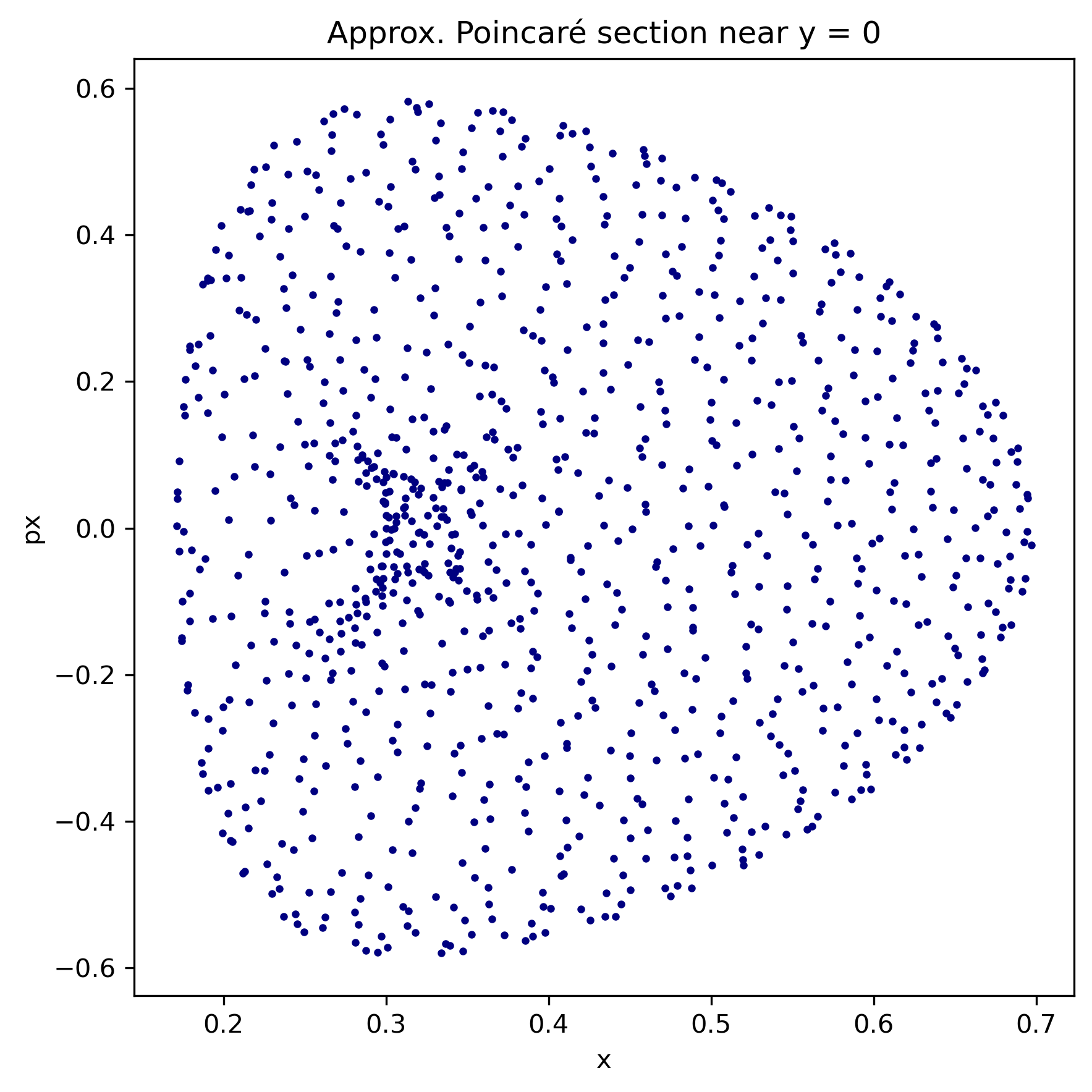}

\caption{Poincaré surface of section obtained by sampling the trajectory at \(y=0\) with \(\dot{y}>0\). The section points remain bounded and display a structured distribution, indicating a mixed dynamical regime with remnants of regular motion and nonlinear perturbations from the ghost-canonical coupling.
$number ~of ~near-section ~points: 1000, ~
x ~ range: 0.17117803268864362  ~ 0.6970359432165766, ~
p_x ~ range: -0.57995466714223  ~ 0.5825208410229082$
}
    \label{fig:Poincare Surface}
\end{figure}

As shown in Fig.~\ref{fig:Poincare Surface}, the points remain confined within a finite region and form a structured pattern rather than uniformly filling the plane. This indicates that the motion is neither purely periodic nor fully ergodic. Instead, the system exhibits a mixed phase-space structure, where remnants of regular motion coexist with irregular trajectories.

The Poincar\'e section therefore complements the phase-space projections by demonstrating that the dynamics are  bounded while retaining nontrivial internal structure.

\subsection{Running Maximal Lyapunov Exponent}
\label{subsec:lyapunov}

The running maximal Lyapunov exponent exhibits a clear transient behavior at early times, including an initial overshoot associated with the finite-time nature of the estimate and the local stretching properties of the orbit in Fig.~\ref{fig:lyapunov}. After this transient, the exponent decreases and approaches a small positive value, indicating that nearby trajectories diverge exponentially on average while the motion remains confined within a bounded region of phase space. This behavior is consistent with weak chaos or stable chaos, in which sensitivity to initial conditions coexists with long-time dynamical confinement.

\subsection{Stable Chaos Regime in the $(\alpha,\varepsilon)$ Plane}
\label{subsec:stable_chaos}

\begin{figure}[htbp]
    \centering
    \includegraphics[width=0.7\textwidth]{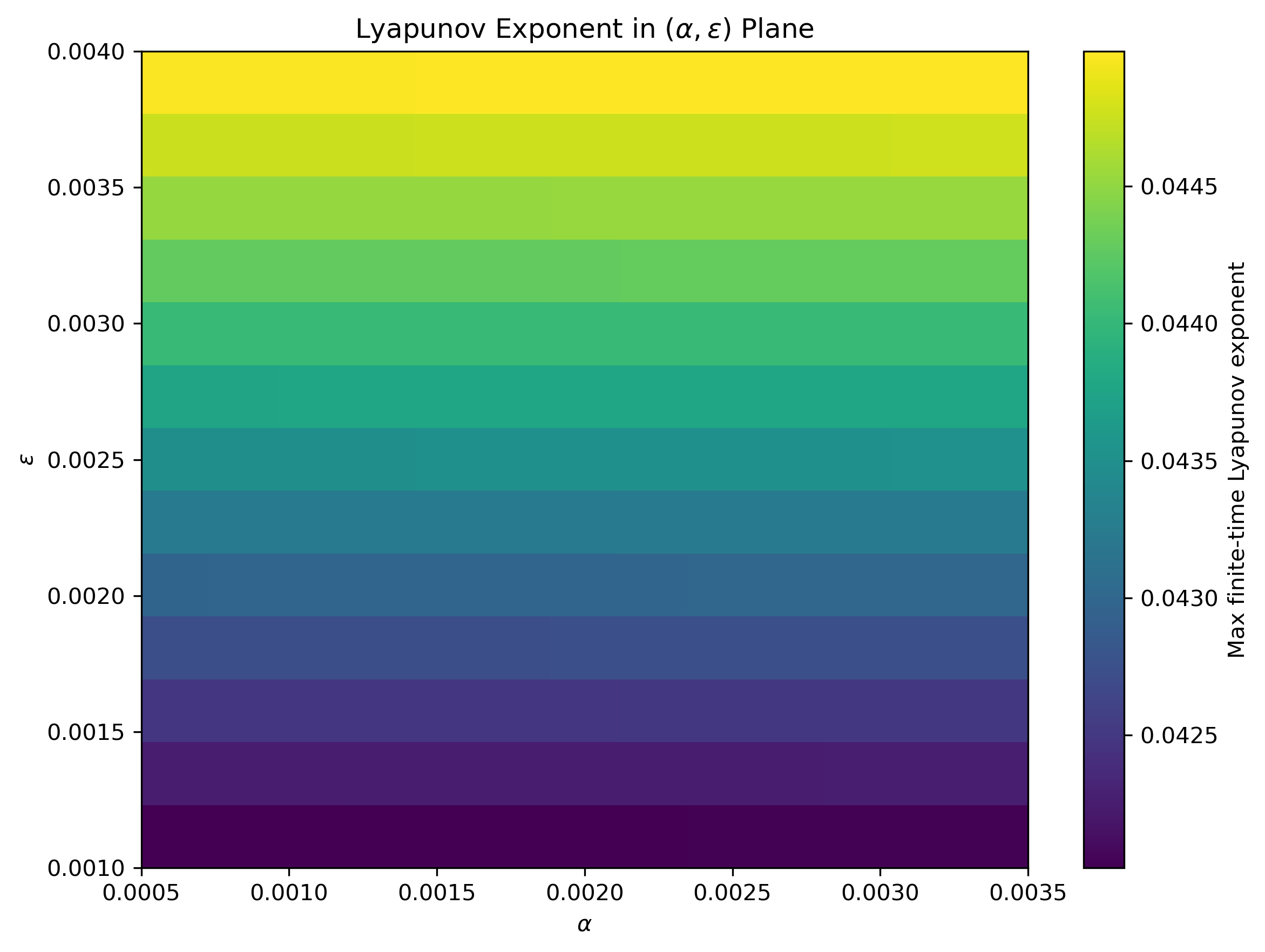
}

\caption{Lyapunov exponent in the $(\alpha,\varepsilon)$ parameter plane, shown as a color-coded heatmap of the maximal finite-time Lyapunov exponent $\lambda_{\max}$. The smooth variation of color across the grid indicates how the strength of sensitivity to initial conditions changes with the perturbation parameters, with lower values corresponding to weaker chaos and higher values indicating stronger chaotic behavior within the explored parameter range.}
    \label{fig:heatmap}
\end{figure}

A central result of this work is the identification of a stable chaos regime in the $(\alpha,\varepsilon)$ parameter plane, in which weak chaos coexists with global boundedness. As illustrated in Fig.~\ref{fig:heatmap}, the maximal finite-time Lyapunov exponent varies smoothly across the parameter domain, indicating that the degree of sensitivity to initial conditions is controlled primarily by the relative strength of the nonlinear perturbation term $\varepsilon x^2y^2$ and the confining contribution $\alpha V_c$.

This regime is characterized by two simultaneous properties:
\begin{itemize}
\item \textbf{Local sensitivity:} The system exhibits a positive maximal Lyapunov exponent, $\lambda_{\max} > 0$, showing exponential separation of nearby trajectories.
\item \textbf{Global boundedness:} Despite this sensitivity, the trajectories remain confined within a finite region of phase space over the explored integration times, with no runaway escape.
\end{itemize}

The structure observed in the $(\alpha,\varepsilon)$ plane reflects the competition between destabilization and confinement. The quartic coupling $\varepsilon x^2y^2$ enhances nonlinear mixing between the canonical and ghost sectors and promotes chaotic behavior, while the saturating potential $\alpha V_c$ provides a soft confining mechanism that suppresses unbounded growth. In this sense, $\varepsilon$ acts as the main chaos-generating control parameter, whereas $\alpha$ regulates the strength of nonlinear confinement.

The heatmap shows that larger values of $\varepsilon$ generally correspond to stronger chaoticity, as indicated by the increase in $\lambda_{\max}$ across the vertical direction of the diagram. By contrast, varying $\alpha$ within the explored range produces a weaker modulation of the exponent, suggesting that in this parameter window the global stability mechanism remains effective and the system stays in a bounded-chaos regime throughout most of the scanned domain.

We therefore define the stable chaos regime as the region in parameter space where trajectories remain bounded while the maximal finite-time Lyapunov exponent remains positive but moderate. In the present scan, this regime occupies the full plotted $(\alpha,\varepsilon)$ domain, with the color gradient indicating a gradual transition from weaker to stronger chaoticity rather than an abrupt loss of stability. This behavior is consistent with a dynamically confined Hamiltonian system in which nonlinear coupling enhances sensitivity, but the confining potential prevents escape.

Finally, the present Hamiltonian is not Liouville integrable, since no second independent global integral of motion in involution is known beyond the total energy. The nonlinear interaction $V_I$, the quartic coupling $\varepsilon x^2y^2$, and the saturating potential $V_c$ destroy separability and generate mixed phase-space behavior. Accordingly, the system should be interpreted as a non-integrable Hamiltonian model exhibiting bounded chaotic motion rather than an exactly solvable integrable system.

\section{Conclusion }\label{sec:conclusion}

In this work, we have performed a detailed numerical and qualitative analysis of a nonlinear Hamiltonian system containing a ghost degree of freedom coupled to a canonical sector. By introducing an exponential $V_c$ potential, we examined whether the pathological instabilities commonly associated with ghost dynamics such as unbounded energy transfer and runaway trajectories can be mitigated within a controlled classical framework.

We allow the ghost degree of freedom to persist but demonstrate that non-polynomial interactions and saturating potentials can dynamically confine the system within a bounded chaotic attractor, preventing the catastrophic runaway typically predicted by the Ostrogradsky theorem.

Our results provide clear numerical evidence that the addition of nonlinear $V_c$ substantially alters the dynamical structure of the system. In particular, for representative parameter values $ (e.g., 
\varepsilon = 0.002 , \alpha=0.001) $, the phase-space trajectories remain bounded over long integration times, indicating that the exponential potential effectively suppresses the runaway behavior typically expected in ghost-dominated systems. This boundedness is consistently observed across the examined initial conditions and is further supported by the structure of the Poincaré sections, which reveal that trajectories are confined to a finite region of phase space.

At the same time, the computation of the maximal Lyapunov exponent yields a positive value $ (
\lambda_
{max}
\approx
 0.01510432528796746
)$, demonstrating that the system exhibits sensitive dependence on initial conditions. The coexistence of bounded motion with positive Lyapunov exponents indicates a nontrivial dynamical regime in which chaos develops without leading to instability in the sense of phase-space escape. This behavior is further corroborated by the mixed structure observed in the Poincaré surfaces of section, where scattered points associated with chaotic trajectories coexist with regular, closed invariant curves corresponding to KAM tori. Such a mixed phase-space structure is characteristic of nonlinear Hamiltonian systems near the transition to chaos, but its emergence in the presence of a ghost sector is particularly noteworthy.

These observations suggest the existence of a regime realized within a finite region of the parameter space 
$(\alpha, \varepsilon)$ in which ghost-induced instabilities are dynamically controlled by nonlinear effects. While our analysis is primarily based on selected parameter values, the persistence of bounded chaotic trajectories under variation of initial conditions indicates that this behavior is not accidental but rather reflects an underlying structural property of the system. The exponential $V_c$ term plays a central role in this mechanism by effectively introducing a restoring force that counteracts the indefinite kinetic contribution of the ghost degree of freedom.

Thus, our findings provide an example in which the presence of a negative-energy sector does not inevitably lead to catastrophic instability at the classical level. Instead, the interplay between nonlinearity and $V_c$ can produce a dynamically rich phase-space structure characterized by bounded yet chaotic motion. Although caution is required in extrapolating these results, this suggests that ghost-like degrees of freedom may admit consistent dynamical realizations under suitable conditions, particularly in effective models where nonlinear interactions are significant. 

\vspace{50mm}

\bibliographystyle{}
\bibliography{main}

\vspace{100mm}

\end{document}